\documentclass[final,3p,times]{elsarticle}

\usepackage{amssymb,amsmath,bm,gensymb}
\usepackage{graphicx,float,subfig}
\usepackage{lineno}
\usepackage{hyperref}

\hypersetup{
    colorlinks=true,
    linkcolor=black,
    citecolor=black,
    filecolor=black,
    urlcolor=black,
}

\biboptions{sort&compress}

\journal{}

\begin{document}

\begin{frontmatter}

\title{Sparse metapiles for shear wave attenuation in half-spaces}
\author[add1]{P. Celli\corref{cor1}}
\ead{paolo.celli@stonybrook.edu}
\author[add2,add3]{I. Nunzi}
\author[add4]{A. Calabrese}
\author[add3]{S. Lenci}
\author[add2]{C. Daraio}
\cortext[cor1]{Corresponding authors}

\address[add1]{Department of Civil Engineering, Stony Brook University, Stony Brook, NY 11794, USA}
\address[add2]{Division of Engineering and Applied Science, California Institute of Technology, Pasadena, CA 91125, USA}
\address[add3]{Department of Civil, Building Engineering, and Architecture, Polytechnic University of Marche, 60131 Ancona, Italy}
\address[add4]{Civil Engineering \& Construction Engineering Management, California State University, Long Beach, CA 90815, USA}

\begin{abstract}

We show that shear waves traveling towards the surface of a half-space medium can be attenuated via buried one-dimensional arrays of resonators---here called \emph{metapiles}---arranged according to sparse patterns around a site to be isolated. Our focus is on shear waves approaching the surface along a direction perpendicular to the surface itself. First, we illustrate the behavior of metapiles, both experimentally and numerically, using 3D printed resonators embedded in an acrylic plate. Then, via numerical simulations, we extend this idea to the case study of an idealized soil half-space, and elucidate the influence of various design parameters on wave attenuation. Results of this work demonstrate that significant wave attenuation can be achieved by installing sparse resonating piles \emph{around} a selected site on the free surface of the medium, rather than placing resonators directly \emph{underneath} that same site. This work might have implications in metamaterial-based wave attenuation applications across scales.

\end{abstract}


\end{frontmatter}

% \linenumbers

\section{Introduction}
\label{s:intro}

% METAMATERIALS
In the context of vibration attenuation and elastic wave control, metamaterials are mechanical systems featuring a wave-carrying medium decorated with arrays of resonating units. Effectively, each resonator acts as a tuned mass damper. When tuned all at the same frequency, these resonators give way to frequency regions of strong wave attenuation called locally-resonant bandgaps~\cite{Liu2000, Huang2009, Hussein2014}. What makes metamaterials appealing from an application standpoint is their capability of attenuating waves with wavelengths much larger than the size of the resonators or their spacing~\cite{Oudich2011, Rupin2014, Celli2015}. A wide variety of elastic media across different length-scales can be turned into metamaterial systems by adding or embedding resonating units, arranged according to desired periodic or non-periodic patterns; examples are metamaterial bars and beams~\cite{Bergamini2014, Attarzadeh2020}, plates~\cite{Oudich2011, Rupin2014}, solids~\cite{Liu2000, Bilal2018} and half-spaces~\cite{Garova1999, Schwan2013}. In addition, the resonating units can be engineered to interact with all types of waves propagating in these media, from flexural to longitudinal, shear and surface-type. In this work, we are interested in attenuating waves in half-spaces.

Depending on the location and nature of the excitation, several types of waves can develop in a half-space. If the source is on or near the surface, surface waves (e.g., of the Rayleigh type) will travel through the medium. Metamaterial systems that can interact with these waves feature above-surface or sub-surface resonators~\cite{Garova1999, Palermo2016, Colombi2016, colombi2016wedge, Miniaci2016, Colquitt2017, Muhammad2019, Palermo2019}. When the source is far below the surface, the waves that travel in the medium are predominantly longitudinal- and shear-polarized bulk waves. If the goal is to shield a location on the surface from these waves, the most logical metamaterial design would feature a number of resonating units located directly below the target location~\cite{Achaoui2016}. In the seismic realm, where this problem is most relevant, this amounts to creating metamaterial-like foundations~\cite{Mitchell2014, Lasalandra2017, Casablanca2018, Basone2019, Colombi2020}.

\begin{figure}[!htb]
\centering
\includegraphics[scale=1.1]{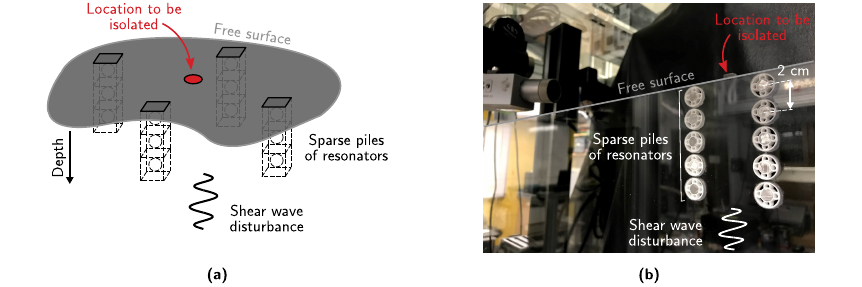}
\caption{The concept of metapiles. (a) Sketch that illustrates the proposed idea of attenuating shear waves with sparse metapiles. (b) Experimental setup for proof of concept. To provide an idea of its scale, we highlight the distance between the centers of two resonators.}
\label{f:intro}
\end{figure}

Here, we are interested in the attenuation of shear waves that are approaching the surface of a half-space from the depth~\cite{Finocchio2014,  Sun2020}. %The study of vertically propagating shear waves is particularly relevant, since they are responsible for strong shaking, and have the potential to inflict the most damage to buildings and their contents. Our work is especially motivated by the fact that metafoundations are \emph{invasive}, in the sense that they cannot be easily applied \emph{a posteriori} to existing structures, and have therefore limited retrofitting potential. In other words, metafoundations inherently require the resonating units to be placed directly below the site or structure to be isolated. 
In particular, we propose the concept of \emph{metapiles}: one-dimensional arrays of resonators buried near the half-space surface and located around -- rather than underneath -- a target location to be isolated. This concept is illustrated in the schematic of Fig.~\ref{f:intro}(a). 
We demonstrate that, when properly designed, sparse arrangements of metapiles allow to significantly attenuate shear waves that impinge on the target location from the depth direction. The key idea behind our concept is that, owing to their subwavelength response, metapiles need not be adjacent to produce wave attenuation effects. In other words, by engineering the distance between piles to be narrower than the wavelength, we ensure that waves cannot be transmitted along those paths between piles and are instead attenuated. In a way, the behavior of metapiles is opposite to that of resonant waveguides designed for wave transmission~\cite{Khelif2004, Oudich2010, Celli2015}, which feature instead paths that are comparable in size to the wavelength.

We first study this concept and the underlying physics, by embedding 3D printed arrays of resonators in acrylic plates, as illustrated in Fig.~\ref{f:intro}(b). Numerical results on this system are validated via experiments. Then, we extend our idea to the case study of a semi-infinite soil medium and resort to numerical simulations, to quantify the performance of metapiles and the influence of various design parameters, such as the characteristics of the resonators, the number of resonators in a pile, the number of piles and the distance between piles. We demonstrate that even a small number of piles and a limited number of resonators can have significant effects on wave attenuation. These considerations are here validated assuming linear elasticity and small-amplitude waves. In the future, the performance of metapiles should be further validated in more realistic scenarios, involving soil-structure interactions and broadband, transient signals. %Since our idea is based on placing resonators around a site to be protected it has the potential to make metamaterial-inspired wave control more applicable as a retrofitting strategy to isolate infrastructure from ground-borne vibrations.

This work is organized as follows. In Section~\ref{s:table}, we show our idea via numerical simulations and experiments on a tabletop model. In Section~\ref{s:seismic}, we translate this idea to wave attenuation in an elastic half-space representing a soil medium. Conclusions and future outlook are discussed in Section~\ref{s:conc}.

%This is a new concept in structural control with metamaterials as studies available in literature have focused on the effects of metamaterials to control the motion of a point when resonators are installed directly below the control node. It is clear that in real applications of these systems, particularly to existing structures, the installation of resonators directly below the control node would be disruptive and costly. As part of this research work we are focusing on determining the effects of metapiles when these are positioned away from the control node. The scheme of Fig. 1a, shows an easy to build configuration that could be used to control existing structures without any modification to their foundations. The implementation of the proposed system is much simpler than that of conventional metamaterials which are generally positioned directly  below the control node  (e.g., [] [][]).
%Explain in detail what we do.
%The way our system works is the opposite of a waveguide. Waveguides in metamaterials can be designed by opening paths of defects and making sure that the path is as large as the wavelength involved. In this case, by placing resonators apart from each other but at a distance that is smaller than the wavelength, we can guarantee that waves do not insinuate in between those arrays. 
%Our work provides a first attempt at quantifying, albeit for specific scenarios, the performance of metapiles with respect to infinitely-extending metamaterial arrays. 

\section{Proof of concept}
\label{s:table}

To understand the physics behind the behavior of our metapiles and study their performance, we develop a two-dimensional experimental setup. The centerpiece of the setup, illustrated in Fig.~\ref{f:intro}(b), is a large acrylic pate in which we embed metapiles, here represented by arrays of 3D printed resonators (ca. 1.8 cm in diameter). As discussed in this section, the design is guided by limitations imposed by the maximum size of the plate and by the minimum feature size allowed by our 3D printer. Leveraging this setup and finite element (FE) simulations performed in COMSOL Multiphysics, we demonstrate the wave attenuation performance of metapiles in an idealized setting, and draw preliminary information on their behavior. 

%We consider a 2D analog to the half-space problem; our goal is to design a setup for metapile testing. Before that, we need to design resonators and make sure that they have shear wave bandgaps. We first design and test the shear wave control performance of composite 3D printed resonators, and then we use these resonators to create metapiles in a 2D acrylic plate. We use numerical simulations to verify our experimental results and to extrapolate them to other scenarios.

\subsection{Resonator design, fabrication and testing}
\label{s:1D}

%Factors to keep into account for the design: (1) shear wavelength in acrylic; (2) minimum 3D printable thickness; (3) k, m ratio in designing the resonator, keeping into account that m has to be large enough for the bandgap to be large; (4) overall resonator size with respect to the size of the plate.

%Describe choice of resonators, materials and fabrication methods. Describe model (3D)

Our experimental setup is based on the idea that an acrylic plate can be representative of a semi-infinite medium. This choice is inspired by our previous work on surface wave control~\cite{Palermo2019}. We create metapiles by carving through holes on the acrylic plate and filling them with composite resonators that feature polymeric springs and metallic masses. Such resonators are common in the metamaterials literature~\cite{Liu2000, Bonanomi2015, Matlack2016, Barnhart2019}. In particular, our objective is to design these resonators for shear wave attenuation. As a first step, we set bounds on acceptable resonant frequencies. Considering the standard properties of polymethyl methacrylate (PMMA), Young's modulus $E = 5.5\,\mathrm{GPa}$, Poisson’s ratio $\nu=0.35$ and density $\rho=1190\,\mathrm{kg\,m^{-3}}$, we can readily calculate the wave speed for shear waves: $v_s=\sqrt{E/(2\rho+2\rho\nu)}\approx1300\,\mathrm{ms^{-1}}$. Assuming nondispersive propagation of shear waves in the plate, the wavelength as a function of frequency $f$ is: $\lambda_s=v_s/f$. Since the size of the plate is limited to a maximum of $1219\times610\,\mathrm{mm}$, and since this size needs to be large enough to accommodate at least a couple wavelengths along the shortest dimension, our frequency of operation has a lower bound of approximately $4\,\mathrm{kHz}$. %The upper bound comes from considerations on the size of the resonators, and the choice depends on the specific resonator design. 

\begin{figure}[!htb]
\centering
\includegraphics[scale=1.1]{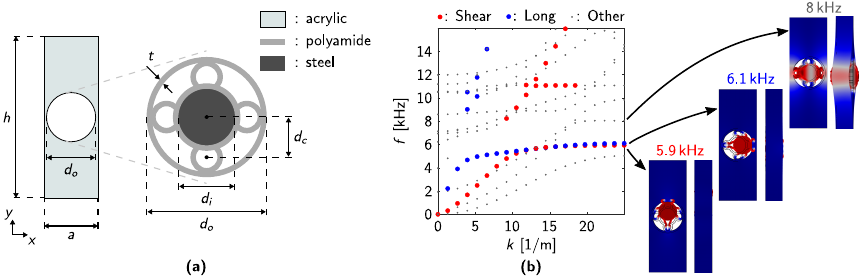}
\caption{3D printed resonator design. (a) Unit cell for the one-dimensional tests, with its relevant dimensions. Its out-of-plane thickness is $b=6.35\,\mathrm{mm}$. Bloch boundary conditions are applied to the long edges, and free boundary conditions to the short edges. The inset illustrates the polyamide-steel resonator. (b) Numerical dispersion relation of the unit cell in (a). The insets illustrate three distinct mode shapes (from lowest to highest frequency: shear, longitudinal and out-of-plane) at $k=\pi/a$.}
\label{f:1D}
\end{figure}

There are two ways of making polymer-metal composite resonators. One way relies on surrounding the metallic mass with a soft elastomeric layer~\cite{Liu2000}; this design is known to feature large damping~\cite{Barnhart2019}. The other avenue, chosen in this work, features 3D printed compliant springs made of a stiff elastomer~\cite{Bonanomi2015, Matlack2016}; this combination is a better choice if the damping within a resonator needs to be minimized. In this work, the 3D printed springs are made of Shapeways' polyamide (PA2200) and fabricated via selective laser sintering. The properties of this material are: Young's modulus $E_p = 1.7\,\mathrm{GPa}$, Poisson’s ratio $\nu_p=0.34$ and density $\rho_p=930\,\mathrm{kg\,m^{-3}}$~\cite{Pajunen2019}. %After testing a few different geometries, we choose the circular resonator 
Our resonator of choice is illustrated in Fig.~\ref{f:1D}(a), and it features a circular polyamide casing with circular springs and a heavy mass at its center. The geometry of the resonator is chosen via trial-and-error, with the objective of obtaining a resonance in the 6\,kHz range.
The thickness of the polyamide walls is $t=0.85\,\mathrm{mm}$, the total diameter of the casing is $d_o=18\,\mathrm{mm}$, the diameter of the heavy mass is $d_i=8.4\,\mathrm{mm}$ and the distance between the center of the heavy mass and the center of one of the circular springs is $d_c=5.7\,\mathrm{mm}$. For the heavy metallic mass, we choose Grade 304 Stainless Steel, whose nominal properties are: Young's modulus $E_s = 193\,\mathrm{GPa}$, Poisson’s ratio $\nu_s=0.27$ and density $\rho_s=8000\,\mathrm{kg\,m^{-3}}$. 

To test this design, we consider as unit cell a strip of acrylic of height $h=60\,\mathrm{mm}$, width $a=20\,\mathrm{mm}$ and out-of-plane thickness $b=6.35\,\mathrm{mm}$, that features a circular hole of diameter $d_o$. The resonator is press-fit into the acrylic and the two parts are bonded via cyanoacrylate. This unit cell is tessellated along the $x$ direction as to form a 1D array. To simulate wave propagation in an infinite array, we use finite element simulations with Bloch periodic boundary conditions. The $h$ dimension of the strip, perpendicular to the direction of wave traveling, is chosen to be significantly larger than $a$ to simulate wave speeds that are comparable to those we expect when the resonators are embedded in a large plate and are not in proximity of the plate's boundaries. The result of this analysis is the band diagram shown in Fig~\ref{f:1D}(b), where each circular marker is automatically color-coded depending on the characteristics of the corresponding mode shape; this allows to identify specific modes of wave propagation. The blue markers correspond to a longitudinal mode, that features a bandgap starting at $6.1\,\mathrm{kHz}$, where the mode flattens; as highlighted by the corresponding mode shape at the edge of the Brillouin zone (corresponding to the maximum $k$ value), this gap is caused by resonant dynamics of the composite resonator. The red markers represent a shear mode, that also features a resonant bandgap starting at $5.9\,\mathrm{kHz}$. This is the mode we are mostly interested in. All gray points in Fig~\ref{f:1D}(b) correspond to mixed modes. Of interest to our discussion is the mode featuring out-of-plane resonant dynamics at $8\,\mathrm{kHz}$.

\begin{figure}[!htb]
\centering
\includegraphics[scale=1.1]{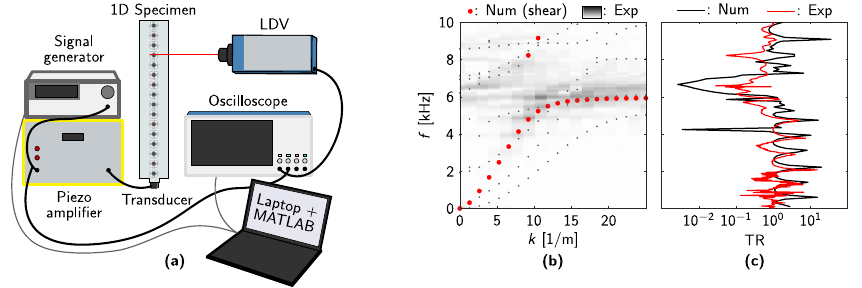}
\caption{Experiments on 1D specimens. (a) Experimental setup to test a one-dimensional array of resonators embedded in acrylic. (b) Numerical (circular markers) and experimentally-reconstructed (colormap) dispersion relations. (c) Numerical and experimental transmission curves.}
\label{f:1Dexp}
\end{figure}

Now that we have evidence of the presence of a shear wave bandgap taking place at acceptable frequencies, we can validate these predictions experimentally using a finite strip of 18 resonators. Our experimental setup is illustrated in Fig.~\ref{f:1Dexp}(a).
The strip is glued to a piezoelectric actuator capable of generating shear waves (Panametrics Videoscan V150-RM), that imparts a wideband Ricker signal centered at $10\,\mathrm{kHz}$. The signal is created in MATLAB, fed to a signal generator (Agilent 33220A), and amplified by a piezo amplifier. To measure the shear component of the traveling wave at various points of the structure, we place a Laser Vibrometer (Polytec) on a motorized linear stage, and program the stage so that the laser can acquire data at each resonator and, in particular, at points where the resonator wall is directly in contact with the acrylic material. Velocity signals are then recorded by an oscilloscope (Tektronix DPO3034) and postprocessed in MATLAB. To provide a complete characterization of the measured mode of wave propagation, we take the data at all measurement points and use it to reconstruct a dispersion relation for the medium via a 2D-Discrete Fourier Transform of the space-time data we obtain. The reconstructed dispersion is shown as a colormap in Fig.~\ref{f:1Dexp}(b). We can see that the dark regions match the markers corresponding to the numerical dispersion relation (same markers as in Fig.~\ref{f:1D}(b)), especially around the point where the lower shear branch flattens (near resonance). Predicting the full extent of the bandgap from this plot is more challenging. To better extract this information, we consider only the measured data at the input (near the transducer), and at the output (the point of the specimen that is further away from the transducer). We then plot the transmission (TR, output velocity divided by input velocity), and we compare it to a numerical prediction obtained via harmonic analysis in COMSOL. Note that no damping is used in these simulations. This comparison is illustrated in Fig.~\ref{f:1Dexp}(c). We can see that the numerical model captures the experimental response for a wide range of frequencies, as highlighted by the proximity of peaks in the two sets of results. Additionally, both numerical and experimental results show a dip in the transmission between 6 and $7.3\,\mathrm{kHz}$, which can be ascribed to the bandgap.

\subsection{Experiments on shear wave attenuation via metapiles}

\begin{figure}[!htb]
\centering
\includegraphics[scale=1.1]{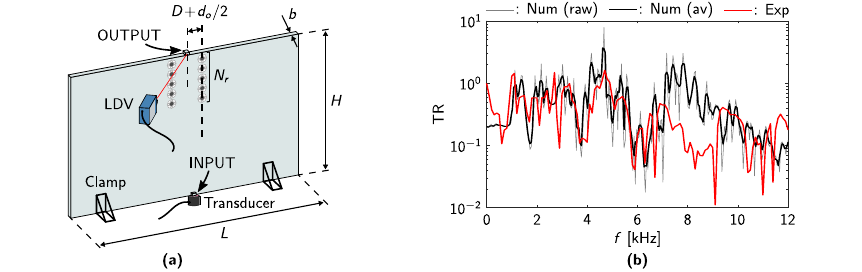}
\caption{Experiments on shear wave attenuation via metapiles. (a) Experimental setup, featuring a large acrylic plate with two metapiles. (b) Numerical and experimental transmission curves. The averaged numerical curve is obtained from a moving average of the raw data. The experimental results, due to the limitations of our setup, account for both in-plane and out-of-plane motion.}
\label{f:2D}
\end{figure}

After validating the behavior of our composite resonators, we can proceed to provide an experimental demonstration of the attenuation behavior of metapiles embedded in a larger acrylic domain. While this idea could be extended to 3D domains, we choose to validate it for a 2D medium for simplicity. A schematic of our experimental setup, with all its relevant dimensions, is illustrated in Fig.~\ref{f:2D}(a); recall that a photo of this same setup is shown in Fig.~\ref{f:intro}(b).
The acrylic domain has width $L=1219\,\mathrm{mm}$, height $H=610\,\mathrm{mm}$ and thickness $b=6.35\,\mathrm{mm}$. Two regions of the bottom edge of the plate, located 30\,cm away from the vertical edges, are clamped to an optical table via angle brackets. The same transducer used for the 1D experiments is also glued at the center point of the bottom edge of the plate, as to simulate a source of shear waves. Near the transducer, we also attach a small acrylic block onto the plate. This is needed to provide a measurement point for in-plane shear waves that is accessible to the laser vibrometer . This measurement point is used to record an input signal. At the top edge of the plate, representing the free surface of our domain, we also glue an acrylic block that is used to record the output response. To properly record in-plane shear waves, the laser should be parallel to the plate itself (i.e., it should be parallel to the direction of vibration that needs to be measured). However, this is not possible due to space limitations in our setup. As a consequence, the laser is oriented at a $\approx30^{\mathrm{o}}$ angle with respect to the plate and is therefore bound to record some out-of-plane dynamics together with the desired in-plane vibration. In order to create and test metapiles, we carve holes in the acrylic plate by means of a CNC router, and press-fit composite resonators identical to those introduced in Section~\ref{s:1D}. Resonators are located right below the top edge (the center of the first resonator is located at $a/2$ from the top edge), and the edge of the resonator of each pile is located at a distance $D$ from the vertical mid-line of the plate, so that the center of the resonator is $D+d_o/2$ distant from that same point. For our experiment, given $d_o=18\,\mathrm{mm}$, we choose $D=21 \,\mathrm{mm}$. This $D$ value is chosen so that the distance between piles $2D=42\,\mathrm{mm}$ is much smaller than the wavelength we expect at the resonance frequency of the resonators, $223\,\mathrm{mm}$. Our specimen features $N_r=5$ resonators for each pile. This setup is replicated in COMSOL, where we perform a harmonic analysis to also determine the transmission of this medium to in-plane shear waves.

A comparison between experimental and numerical transmission curves is shown in Fig.~\ref{f:2D}(b). Our numerical simulations do not feature any damping. To smoothen-out the numerical frequency response and to make it resemble the response one would obtain with moderate values of damping, we apply a moving average procedure. The response before and after the application of the moving average is shown as gray and black curves in Fig.~\ref{f:2D}(b). The red curve is the experimental transmission curve. We can see that there is good qualitative agreement between numerics and experiments up until 7 kHz. In particular, we can see that both sets of data capture the small dip around 4 kHz, the amplitude increase around 5 kHz and the large dip that starts right below 6 kHz. Based on our knowledge on the dynamics of these resonators (section~\ref{s:1D}), we understand that this large dip is the onset of the bandgap induced by the metapiles. After 7 kHz, we can see that the numerical response increases again, while the experimental one experiences a second large dip. We claim that this second dip is due to undesired out-of plane dynamics of the plate that are picked up due to the inclination of the laser. This conjecture is corroborated by the fact that the band diagram in Fig.~\ref{f:1D}(b) shows that our resonators present an out-of-plane resonance in the neighborhood of 8 kHz, the frequency at which we see the second dip in Fig.~\ref{f:2D}(b). In conclusion, this preliminary experiment demonstrates that shear wave attenuation in a half-space can take place even if we use spaced-out arrays of resonators. Next, we will use numerical simulations to better understand this attenuation phenomenon.

\subsection{Numerical generalization and parametric study}

\begin{figure}[!htb]
\centering
\includegraphics[scale=1.1]{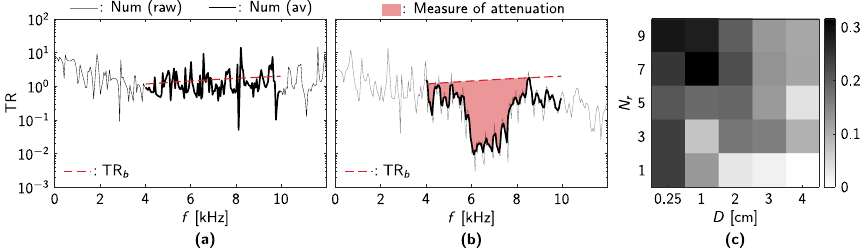}
\caption{Numerical parametric study. (a) Transmission curve of the bare plate (no resonators). TR$_b$ is the baseline transmission in the 4--10$\,\mathrm{kHz}$ range of interest, obtained by fitting the transmission with a first order polynomial. (b) Extraction of the attenuation measure from the transmission of a configuration with resonators. This specific configuration, featuring 20 adjacent piles with 5 resonators per pile, is used as reference to benchmark attenuation performance. The shaded area is obtained by intersecting the transmission with the TR$_b$ line. (c) Evolution of the attenuation performance with $N_r$, the number of resonators in a metapile, and $D$, the distance between the edge of one metapile and the output measurement point. The attenuation measure for each case is normalized by the area of the attenuation region in the benchmark case (b).}
\label{f:num}
\end{figure}

In order to probe the effects of various metapile parameters on the attenuation performance, it would be necessary to test many different spatial resonator configurations. This is very impractical to do experimentally. However, since we have validated our numerical simulations with experimental results for a specific choice of parameters, we now resort to numerical simulations to perform a limited parametric study. In particular, we keep the same identical resonators as in previous sections, and vary $D$, the distance between the edge of a metapile and the location to be shielded, and $N_r$, the number of resonators in each metapile. To compare the performance of various configurations, we define a metric of wave attenuation performance as illustrated in Fig.~\ref{f:num}(a,b).
First, we consider a frequency range around the expected bandgap, here chosen to be from $4\,\mathrm{kHz}$ to $10\,\mathrm{kHz}$. We then compute the transmission of a bare acrylic plate without resonators and extract a baseline curve (linear fit of the response in the desired frequency range). This procedure, shown in Fig.~\ref{f:num}(a), is done to account for the slope of the response curve in the range of interest. When we consider a system with resonators, we first smoothen out the numerical transmission with a moving average filter. Then, we consider as bandgap a continuous region that includes the expected resonance $6\,\mathrm{kHz}$ and that remains below the baseline. The area of that region is our measure of attenuation.

\begin{figure}[!htb]
\centering
\includegraphics[scale=1.1]{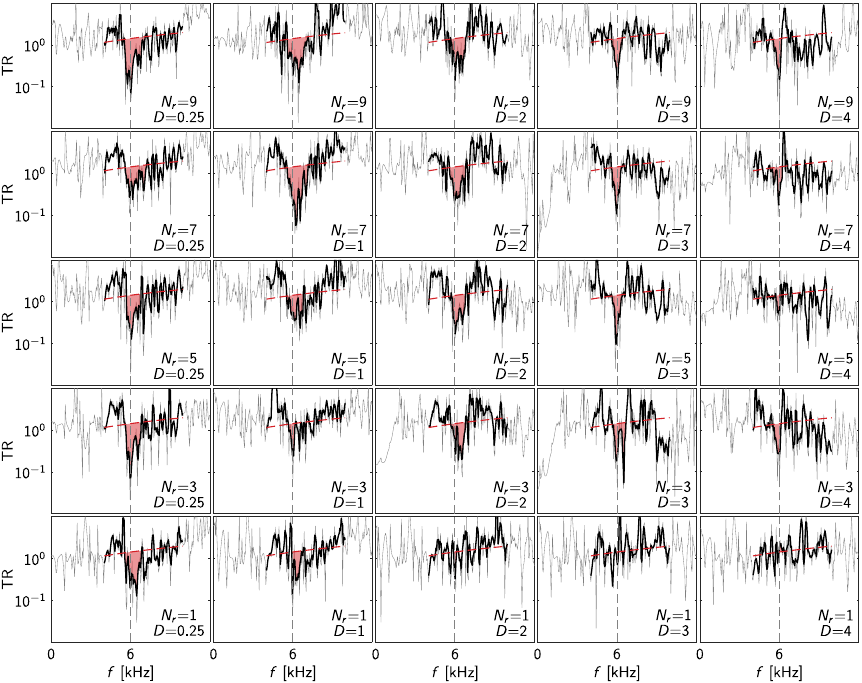}
\caption{Complete map of the transmission results that yield the parametric colormap in Fig.~\ref{f:num}(c). All simulations feature a line wave source.}
\label{f:numall}
\end{figure}

For each metapile configuration of interest, we extract the attenuation area and normalize it by the area obtained for a compact array of $20\times5$ resonators located right below the location to be shielded. The transmission plot and attenuation area for this reference configuration are shown in Fig.~\ref{f:num}(b). By considering the normalized areas for combinations of $D=0.25\text{--}4\,\mathrm{cm}$ and $N_r=1\text{--}9$, we can extract information to compile the design map shown in Fig.~\ref{f:num}(c). To better appreciate the extent of the attenuation area for each metapile configuration, one can see all transmission plots in Fig.~\ref{f:numall}.
From this data set, we can conclude that configurations with more resonators in the metapile and with less distance between piles perform better in terms of wave attenuation. In particular, from the map in Fig.~\ref{f:num}(c), we can see that the performance of the best metapile configurations (with low $D$ and high $N_r$) is approximately 30\% of the performance of the reference configuration --  a metafoundation featuring a number of resonators an order of magnitude larger than any metapile configuration. From these plots, we can also appreciate that $D$ has much larger effects than $N_r$, as highlighted by the fact that configurations with metapiles close to each other, but featuring a few resonators, perform better than configurations with many resonators per pile but with far-away piles. This is not surprising since, for $D=4\,\mathrm{cm}$, the distance between piles of $2D=8\,\mathrm{cm}$ is much closer in magnitude to the shear wavelength in the acrylic plate.

\section{Application to a soil half-space}
\label{s:seismic}

%Important segue. Why are we doing this half-space model? Results of previous section do not give us information that is meaningfully scalable and that can be applied at larger scales. Problems are size of the resonators with respect to the wavelength, limited size of the plate, etc. Thus, we consider an elastic half-space in SAP 2000 and simulate the effect of metapiles on shear wave attenuation. We also perform a full parametric study.

The results illustrated in Section~\ref{s:table} give us some preliminary information on the wave attenuation performance of metapiles. However, two aspects make vast parametric studies based on the model used in Section~\ref{s:table} impractical: (1) the size limitations of our experimental setup, that does not allow us to increase the distance between piles without incurring in significant boundary effects; and (2) the fact that our COMSOL models are based on an accurate rendering of each resonator, and are therefore computationally expensive. Thus, we build a new model where resonators are simplified as spring-mass systems connected at single nodes of the half-space. We also take advantage of this new model to concentrate on a case study that more closely resembles a potential application of our system. We therefore consider an elastic half-space with soil-like properties. Using the software SAP 2000, a structural engineering oriented FE platform, we investigate the effects of different layouts of metapiles and their characteristics on shear wave attenuation.

%The aim of the analyses discussed in this section is that of extending the results obtained on a small tabletop models to the elastic half-space. A parametric study was performed in SAP2000 to investigate the effects of different layouts of metapiles on shear wave attenuation. 

\subsection{The numerical model}

%Here we need a description of the numerical model in SAP. Wg=hat properties we chose. Why? References on how the half-space properties were selected. Describe the model and the boundary conditions (Fig.7a). First, talk about response without piles and describe resonances. Peak of interest is 3 kHz. Then, describe what happens when resonators are everywhere on upper 5 layers. Describe the measures of wave attenuation. PA: peak attenuation; EA: effective attenuation; BW: bandwidth. 

\begin{figure}[!htb]
\centering
\includegraphics[scale=1.1]{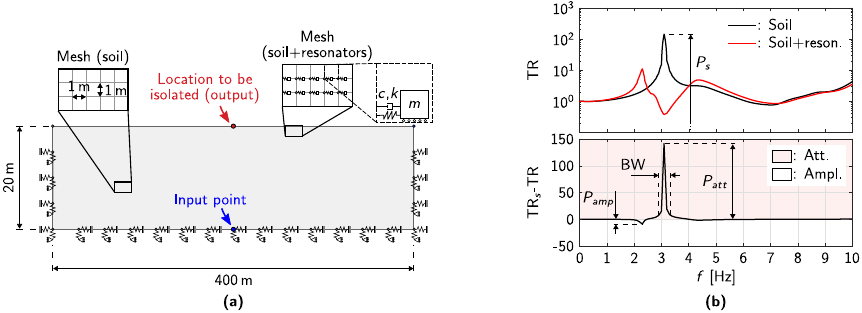}
\caption{(a) Elastic half-space model in SAP2000. (b) Top panel: comparison between the response of the soil and the response of the soil with 5 layers of resonators located below the surface. The resonators are tuned at the first peak of the soil response. $P_{s}$ is the amplitude of the soil response at 3.1\,Hz. Bottom panel: difference between the transmission}
\label{f:sapmod}
\end{figure}

The half-space model, illustrated in Fig.~\ref{f:sapmod}(a) has a base of $400\,\mathrm{m}$, a depth of $20\,\mathrm{m}$ and a mesh of four-nodes $1\,\mathrm{m}\times1\,\mathrm{m}$ elements.
We assume plain strain conditions. The half-space is assumed to be linear elastic, isotropic, and homogeneous. Soil conditions are modeled assuming a density of the material of $1700\,\mathrm{kg\,m^{-3}}$, an elastic modulus of $7.07\,\mathrm{GPa}$ and a Poisson's ratio of 0.3. We consider a damping ratio of 5\%. Frequency domain analyses are performed by applying a horizontal (shear) harmonic base displacement of $0.05\,\mathrm{m}$ to the baseline of the model. Frequencies of excitation are selected in the range $0\text{--}25\,\mathrm{Hz}$. We introduce absorbing boundary conditions using dampers, as proposed by Lysmer and Kuhlemeyer~\cite{lysmer1969finite}. Along the boundaries, we choose damping constants $c_1=1\,\mathrm{Nsm^{-1}}$ and $c_2=0.25\,\mathrm{Nsm^{-1}}$ along the horizontal and vertical directions, respectively, as studies show that this assumption leads to a reasonable wave absorption~\cite{wang2009numerical, vermeer1998plaxis}. To reduce wave reflections at the lateral boundaries of the domain, we also choose a lateral extension of the domain that is more than eight times its height ~\cite{amorosi2009numerical}. The size of the mesh is selected to meet the requirements proposed by Lysmer and Kuhlemeyer, as each element has dimensions much smaller than $\lambda/8$, where $\lambda$ is the wavelength corresponding to the maximum frequency of interest $f$. For the analyses discussed in this work, $\lambda/8=v_s/8f=16\,\mathrm{m}$, being $v_s=396\,\mathrm{ms^{-1}}$ and $f=3.1\,\mathrm{Hz}$. Finally, to perform a transmission analysis, we record the lateral displacement at an input point located at the midpoint of the baseline of the model, and at an output point (also called location to be isolated or control node) located at the midpoint of the upper boundary. For the half-space without resonators, we compare the peaks of the numerical transmission, shown as a black curve in Fig.~\ref{f:sapmod}(b), to the theoretical solution available in the literature~\cite{kramer1996geotechnical}. This comparison yields a perfect match. Our frequency of interest is $3.1\,\mathrm{Hz}$, corresponding to the first shear resonance of the half-space. 

Metapiles are modeled as arrays of resonators located at prescribed nodes of the mesh; thus, we assume that each resonator occupies an area of $1\mathrm{m}\times1\mathrm{m}$. Since we are interested in shear wave attenuation, the resonators are only capable of lateral motion. An example of transmission for a configuration featuring 5 rows of resonators located below the whole upper boundary of the domain is shown as a red line in the top panel of Fig.~\ref{f:sapmod}(b). These resonators are tuned to resonate at $3.1\,\mathrm{Hz}$, acting as a single, tuned mass damper: the original peak of the half-space is replaced by an anti-resonance accompanied by two adjacent and distinct peaks. To quantify the effects of the resonators on the response of the model, we plot the difference between the transmission of the soil and that of the soil with resonators, as shown in the bottom panel of Fig.~\ref{f:sapmod}(b). A difference greater than 0 corresponds to attenuation regions, while a difference smaller than 0 means that the resonators have amplified the response. This plot clearly highlights that attenuation at 3.1\,Hz comes at a cost: the response is amplified at other frequencies. We then define the following metrics of attenuation: (i) the peak attenuation, defined as $P_{att}/P_s$, where $P_{att}$ is the maximum attenuation and $P_{s}$ is the soil response at the resonance peak of $3.1\,\mathrm{Hz}$; (ii) the effective attenuation, which is meant to account for the presence of amplification regions and is defined as $(P_{att}-P_{amp})/P_s$, where $P_{amp}$ is the maximum amplification; (iii) the bandwidth BW, i.e., the frequency range where a peak attenuation greater than 5\% is detected.

\subsection{Parametric study}

\begin{figure}[!htb]
\centering
\includegraphics[scale=1.1]{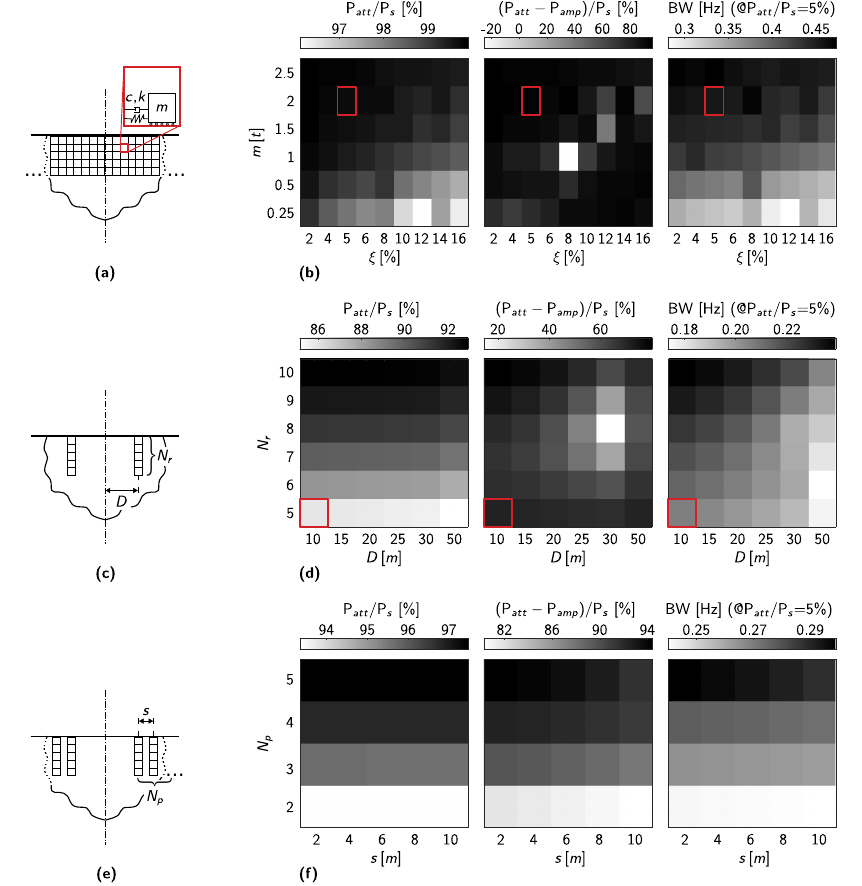}
\caption{Parametric study on the soil half-space model. Various configurations of resonators are here compared as a function of peak attenuation $P_{att}/P_s$, effective attenuation $(P_{att}-P_{amp})/P_s$, and bandwidth BW. (a) Resonator arrangement, featuring 5 rows of resonators homogeneously distributed below the free surface, used to study the influence of the resonators parameters. (b) Design maps obtained from (a), detailing the influence of the resonators' mass $m$ and damping factor $\xi$. (c) Generic metapile configuration with two arrays of resonators that are equidistant from the control node. (d) Design maps obtained from (c), detailing the influence of the distance between metapile arrays and control node $D$, and the influence of $N_r$, the number of resonators in each array, with parameters $m=2\,\mathrm{t}$ and $\xi=5$\% fixed. (e) Configuration with more than two metapiles. (f) Design maps obtained from (e), detailing the influence of the number of metapiles located on the same side of the control node $N_p$, and their distance $s$, with $m=2\,\mathrm{t}$, $\xi=5$\%, $D=10\,\mathrm{m}$, $N_r=5$ fixed.}
\label{f:sapmap}
\end{figure}

We begin our parametric study by investigating the influence of the characteristics of the resonators on the attenuation. This comparison is carried out by considering 5 rows of resonators homogeneously distributed below the free surface. This configuration is illustrated schematically in Fig.~\ref{f:sapmap}(a).
We let the mass of each resonator range from $0.25\,\mathrm{t}$ to $2.5\,\mathrm{t}$, and the equivalent viscous damping factor from 2 to 16\%. We keep the natural frequency of the resonator constant at $3.1\,\mathrm{Hz}$ and therefore vary the stiffness of the resonator to tune its response. From the charts of Fig.~\ref{f:sapmap}(b), it is interesting to observe that the peak attenuation is generally better for higher mass and lower damping. We can see that, in this configuration featuring a large number of resonators, even resonators with low masses and high values of damping yield a $\approx95\%$ attenuation of the peak. Similar trends are observed for the bandwidth, that peaks for large masses at about $0.45\,\mathrm{Hz}$. It is less straight-forward to make meaningful considerations for the effective attenuation; the only interesting feature of this map is the presence of an outlier for $\xi=8\%$ and $m=1\,\mathrm{t}$. For this combination of parameters, the effective attenuation is smaller than 0 and indicating amplification of the signal. This highlights the importance of also considering amplification effects in addition to attenuation -- an aspect that is discussed in other works~\cite{Xiao2020} and whose detailed investigation in the case of metapiles deserves a separate treatment. Since a mass of $2\,\mathrm{t}$ and a damping of 5\% allow to obtain a significant response attenuation, we fix these reasonable parameters in the following analyses. 

To study the effects of varying numbers of resonators $Nr$ in a pile and of their distance from the control node $D$, we build the model illustrated in Fig.~\ref{f:sapmap}(c). Fig.~\ref{f:sapmap}(b) shows that, when resonators are homogeneously distributed below the surface, the peak attenuation for a mass of $2\,\mathrm{t}$ and damping of 5\% amounts to $99\%$. Fig.~\ref{f:sapmap}(d) shows that the peak attenuation we obtain with only two metapiles located at any distance from 10 to 30 m is still significant. This amounts to 86\% for $N_r=5$ and increases up to $97\%$ for $N_r=10$. Not surprisingly, deeper metapiles yield better peak attenuation. As $D$ increases, we can see that the peak attenuation decreases since the distance between piles $2D$ is now approaching $129\,\mathrm{m}$, the wavelength in the soil at $3.1\,\mathrm{Hz}$. Note that values of $D$ larger than 50 would compromise the validity of the model since the piles would be too close to the left- and right- boundaries of our domain. The effective attenuation map yields less intuitive results, and features a minimum for $D=30\,\mathrm{m}$ and $N_r=8$. Finally,  we can see that the bandwidth decreases as we increase $D$ and decrease $N_r$.

We also analyze the influence of the number of adjacent metapiles, by fixing $m=2\,\mathrm{t}$, $\xi=5\%$, $N_r=5$ and $D=10\,\mathrm{m}$. We call $N_p$ the number of adjacent piles per side of the control node, as shown in Fig.~\ref{f:sapmap}(e). We call $s$ the distance between piles. The wave attenuation performance of these configurations is shown in Fig.~\ref{f:sapmap}(f). Increasing $N_p$, while keeping $D$ constant, increases the attenuation performance of the system in terms of peak attenuation, effective attenuation and bandwidth. The distance between piles $s$ does not have a significant influence on the attenuation performance.

\section{Conclusions}

In this article, we have shown the shear wave attenuation properties of metapiles buried in an elastic half-space. First, we demonstrated via numerical simulations and experiments that cm-scale metapiles embedded in an acrylic plate can attenuate waves, when their distance is smaller than the wavelength in the medium of interest. Then, we numerically extended this idea in waves propagating in soil, showing that metapiles have the potential to yield significant wave attenuation with a minimal number of resonators. %Owing to the fact that metapiles can be quite distant from each other, they have potential as a noninvasive means for bulk shear wave attenuation.

Future work might be directed towards the realization of metapiles with spatially-varying distributions of resonance frequencies, to widen the frequency bandwidth of the wave attenuation regime~\cite{Colombi2020}. For seismic applications, it will be important to evaluate the performance of metapiles in real soil media, known to hinder some of the wave attenuation effects exhibited by metabarriers~\cite{Palermo2018, Zaccherini2020}, and consider soil-structure interaction effects~\cite{Sun2020}. Another important aspect that deserves to be investigated is the coupling between metapiles and the structure to be isolated~\cite{Xiao2020}. In particular, the effects of undesired vibration amplification should be carefully considered. 

\label{s:conc}

\section*{Acknowledgments}
We wish to thank Antonio Palermo for fruitful discussions, and Sai Sharan Injeti, Alex Ogren and Semih Taniker for helpful COMSOL-related assistance.

\bibliographystyle{elsarticle-num}
% \bibliography{ref}

\end{document}